\documentclass[aip,preprint,showkeys,author-year]{revtex4-1}

\makeatletter
\let\@fnsymbol\@fnsymbol@latex
\@booleanfalse\altaffilletter@sw
\makeatother

\usepackage[pdftex]{graphicx}
\usepackage{float}
\usepackage{mathptmx}
\usepackage{amsmath}
\usepackage{latexsym}
\usepackage{color,colortbl}
\usepackage[table]{xcolor}
\usepackage{tabularx}
\usepackage{hyperref}
\begin{document}

% \preprint{APS/123-QED}

\title{Modeling Collective Anticipation and Response on Wikipedia}  % Force line breaks with \\
%  Manuscript Title:\\with Forced Linebreak}

\author{Ryota Kobayashi}
\email{r-koba@k.u-tokyo.ac.jp}
\affiliation{The University of Tokyo}
\affiliation{JST PRESTO}

\author{Patrick Gildersleve}
\email{patrick.gildersleve@oii.ox.ac.uk}
\affiliation{University of Oxford}

\author{Takeaki Uno}
\email{uno@nii.jp}
\affiliation{National Institute of Informatics}

\author{Renaud Lambiotte}
\email{renaud.lambiotte@maths.ox.ac.uk}
\affiliation{University of Oxford}

\date{\today}% It is always \today, today,

\begin{abstract}
The dynamics of popularity in online media are driven by a combination of endogenous spreading mechanisms and response to exogenous shocks including news and events. However, little is known about the dependence of temporal patterns of popularity on event-related information, e.g. which types of events trigger long-lasting activity. Here we propose a simple model that describes the dynamics around peaks of popularity by incorporating key features, i.e., the anticipatory growth and the decay of collective attention together with circadian rhythms. The proposed model allows us to develop a new method for predicting the future page view activity and for clustering time series. To validate our methodology, we collect a corpus of page view data from Wikipedia associated to a range of planned events, that are events which we know in advance will have a fixed date in the future, such as elections and sport events. Our methodology is superior to  existing models in both prediction and clustering tasks. Furthermore, restricting to Wikipedia pages associated to association football, we observe that the specific realization of the event, in our case which team wins a match or the type of the match, has a significant effect on the response dynamics after the event. Our work demonstrates the importance of appropriately modeling all phases of collective attention, as well as the connection between temporal patterns of attention and characteristic underlying information of the events they represent.
\end{abstract}

% \pacs{Valid PACS appear here}% PACS, the Physics and Astronomy
\maketitle

%\tableofcontents
%
%
\section{Introduction}
In recent years, many aspects of human activities have become increasingly mediated by digital services, leaving electronic footprints as an invaluable resource revealing the forces driving the structure and dynamics of social systems~\citep{Backstrom2006,crane2008}. A central question of computational social science is to understand the mechanisms by which individuals, taken as groups, exhibit collective behaviours~\citep{Lazer2009}. This relationship is particularly striking when considering the emergence and subsequent decline in popularity, or success, on the web and in social media~\citep{Szabo2010,Goel2010,Bandari2012,Proskurnia2017,Candia2019}. Take the adoption of certain hashtags, instead of others, when confronted with new social phenomena \citep{Lin2013}; or the fact that certain songs become extremely popular while most remain in the dark \citep{salganik2006experimental}. Several studies have looked at large-scale, temporal datasets or performed online experiments in order to identify the mechanisms, or the lack of them, explaining what makes a specific item successful \citep{hofman2017prediction}. In this context, it is now understood that a successful outcome arises due to a complex combination of chance, of multiplicative, cascading effects and of intrinsic quality, in different proportions depending on the system under scrutiny \citep{janosov2019success}. 

More specifically, the emergence of peaks of popularity in online social systems has attracted much attention~\citep{crane2008,Lehmann2012}. 
Peaks emerge due to endogenous forces defined as interactions within social media, or as a response to exogenous shocks (e.g., news and disasters) — and most of the time due to a combination of both factors. 
Peaks may also emerge around planned events, whose date is known ahead of the realisation of the event, where the anticipation of the users  plays an important role to shape the dynamics. Representative examples include political elections or sport events. It is the main focus of this article to model the dynamics of collective attention around such planned events.

%  Knowledge Gap: Reviewer 1 & 3. 
In this study, we investigate the question of how people respond to planned  events in an online setting. 
While several previous studies have focused on the  dynamics of popularity, they have mostly considered specific events, e.g., movie release~\citep{Mestyan2013}, elections~\citep{yasseri2016wikipedia}, and airplane crashes~\citep{garcia2016dynamics,Garcia2017}, or, looking for universal patterns, have not paid attention to the type of events \citep{crane2008,Matsubara2012,Zhao2015,Kobayashi2016,Proskurnia2017}. 
For these reasons, it remains unclear how event-related information (e.g., category and outcome of an event) influences its peaks of popularity. 
Addressing this question is important for different reasons. 
First, it is essential to develop efficient mathematical models of popularity dynamics for the automatic identification of online events ~\citep{Petrovic2010,Becker2012}, early detection of emerging stars and viral content on the Internet~\citep{TaTar2014}, and evaluation of the effect of recommender systems~\citep{Wu2019}.
Second, this understanding of popularity peaks in online systems translates to an insight about the mechanisms of popularity, trends, and fads offline. Previous studies have demonstrated that popularity in online systems is highly correlated with other offline indicators, including stock market volumes \citep{Bordino2012}, 
movie box office success \citep{Mestyan2013}, the number of people infected with influenza \citep{Hickmann2015}, and election results~\citep{yasseri2016wikipedia}. 

%  Research Questions:
Here we focus on a range of planned events in order to explore the following two research questions (RQs):
\begin{description}
    \item[RQ1]	What are the essential characteristics of collective attention dynamics towards planned events?
    \item[RQ2]	How is event-related information (e.g., category and outcome) associated with the dynamics of collective attention? 
\end{description} 
To answer these questions, we concentrate on page views on Wikipedia and collect data for planned events in a wide range of contexts. We model Wikipedia time series by incorporating the anticipation and response to an event, as well as circadian rhythm. 
Importantly, the interpretability of model parameters allows us to quantify more robustly the relationship between the collective attention before and after the event, differentiating between their volume and time scale. Based on this model, we develop a Bayesian method for predicting future page view activity and perform a clustering analysis based on time series of collective attention. 
Furthermore, we restrict our analysis to a data set associated to football matches to investigate how the temporal pattern of collective attention is related to the event outcome. 
\section{Related Work}
Within the field of computational social science, several works have focused on popularity peaks in social media or on the Web. 
In \cite{crane2008}, the authors analyzed Youtube data and identified two classes in the collective attention dynamics, sudden peak with rapid relaxation, associated to exogenous shocks and gradually growing patterns until the peak, followed by a symmetric relaxation, associated with endogenous effects. 
Other works have proposed taxonomies of popularity peaks based on their temporal profiles. 
For instance, in \citet{Yang2011}, the authors considered the time-series of popularity of blog posts and news media articles, which they clustered based on a similarity metric invariant to scaling and shifting. 

In a related study \citep{Lehmann2012}, the authors analyzed hashtag data on Twitter and identified four classes in the collective attention dynamics: 
1) Sudden peak and rapid relaxation, 
2) Significant growth before the peak and symmetric relaxation, 
3) Significant growth before the peak and sudden decay, and  
4) Sudden peak and sudden decay.
Furthermore, they conducted semantic analysis using WordNet semantic lexicon.
The main differences between our work and this state-of-the-art is the fact that we investigate the association between event-related information (e.g., category and outcome) and collective attention dynamics. 
Our interpretable model describes all the four classes identified in \citet{Lehmann2012} and allows us to develop a new method for predicting the future page view activity and clustering time series. Another critical difference is that we focus exclusively on planned events. 
Studies that pay special attention to planned events are more limited. \citet{Becker2012} tackle the task of retrieving online content for planned events across a range of different social media platforms. 
This work focused on developing the event detection algorithm, but did not investigate the relationship between event-related information and collective attention dynamics, as considered here. 

Our work is relevant to the question of how the popularity of fads rises and falls. Fads, defined as cultural objects including phrases, names, and customs that are very popular for a short period, have been an important research topic in the social sciences ~\citep{burgess1933introduction,aguirre1988collective,abrahamson1991managerial}. However, there has only been a small number of works looking at the temporal property of fads (for exceptions, see ~\citet{strang2001search}, \citet{rich2008management}, \citet{berger2009adoption}, and  \citet{denrell2015effect}) compared to the research on the spread of innovation (see references in \citet{rogers2010diffusion}). Fads are considered as two essential processes ~\citep{strang2001search,rich2008management}: the adoption and subsequent abandonment of an object. While most studies on innovation \citep{rogers2010diffusion} focus on adoption, they pay little attention to the abandonment. In general, it is difficult to track the temporal pattern of popularity of real-world phenomena. 
Whilst online reactions to news events are not strictly fads, there are certainly parallels between the dynamics of anticipation / adoption and forgetting / abandonment. This link is particularly pertinent when the computational social science literature typically focuses on post-event response and not the anticipation / adoption phase. 

Another stream of works related to our contribution concerns the modelling and prediction of time series in social media. Within this question, part of the research focuses on the mechanisms that may explain the growth, or the lack of it, of certain items online. In \citet{Lin2013}, for instance, the authors analysed the competition between novel hashtags during the 2012 U.S. presidential debates and investigated mechanisms by which certain hashtag take over. Those mechanisms include preferential attachment and competition driven by the limited amount of attention of users \citep{Weng2012}. 
From a modelling perspective, various time series models have been proposed to predict the  dynamics of popularity in online social systems. 
Important works include \citet{Matsubara2012}, proposing a time series model, SpikeM, that incorporates an exponential rise, power-law decay, and circadian rhythms.  
In \citet{Proskurnia2017}, the authors proposed a time-series model that incorporates reinforcement and  circadian rhythms, allowing to predict the popularity dynamics on thepetitionsite.com. 
While the previous works developed time series models for predicting popularity dynamics, our work additionally exploits a model in order to investigate the relationship between event-related information and popularity dynamics.

Our work also finds connection with the growing research effort on understanding Wikipedia, our main source of data. 
Most importantly for our research, several works have shown that popularity on Wikipedia is strongly related to popularity in other online services. 
User surveys in \citet{singer2017we} show `media' and `current events' as key motivations for browsing Wikipedia (30\% and 13\% of respondents respectively), likely the same driving forces as other online media. 
To compare platforms more directly; there is a strong correlation between the Wikipedia page view activity and Google search activity \citep{Ratkiewicz2010,Yoshida2015}, despite certain differences, e.g., trends on Twitter and Wikipedia are more ephemeral than on Google, both rising and declining rapidly for newly emerging topics in \citet{Althoff2013}. Taken together, these results indicate that Wikipedia page views reflect users' behavior on the Internet in general. 

In addition to the consumption of information, a range of works cover the dynamics of the production of knowledge on Wikipedia. Whilst we do not study the relationship between the production and consumption of information, they are intimately connected. The interaction between different editors on the supply side of information can yield rich dynamics, through both collaborative efforts \citep{keegan2011hot,keegan2012editors,West2012}, and conflicts \citep{yasseri2012dynamics}, frequently in response to current events. In addition, demand for information on Wikipedia in response to current events is correlated with yet typically precedes, or even drives, supply (as measured by page views and the creation of new articles respectively) \citep{ciampaglia2015production}. Thus, understanding the dynamics of collective attention is also key to studying the production of knowledge on Wikipedia and beyond. 
\section{Data}
\noindent
We consider the popularity dynamics of events in 5 categories;  Elections, Sports events,
Association Football matches (also known as soccer and abbreviated to ``Football"),   
Films (with release date), and Holidays. 
In the early stages of this study, we collected various classes of events over a wider range of categories, including Armed conflicts, Arts, Culture, Business, Economy, Disasters, Accidents, and International relations, among other things. We discarded the other categories after deciding to focus on events that are planned in advance and organized on a single day. Although we limit the scope to this type of event in this study, the proposed model could be generalized to the other types of events. 
Data for events from each class was obtained by scraping table data from relevant English Wikipedia summary articles.
This included the event name (with linked Wikipedia article), date, as well as any class specific auxiliary data such as competition winner, or film director. In the case of  football matches, articles for the two teams competing in each match were collected. Events that did not have a corresponding Wikipedia article were removed from analysis. 
Table~\ref{tab:DataStat} summarizes the statistics of this data set\footnote{All datasets and code available at \url{github.com/NII-Kobayashi/Collective-Anticipation-and-Response}.}.
\begin{table*}[t]
\centering
\begin{tabular}{l|c|c|c}
Category            & Summary Articles &   \# events    &   Period    \\  \hline
% Election			&  [2015-18] national electoral calendar	&	342		&	01/07/15\footnotemark[3]-31/12/18		\\
 Election			&  [2015-18] national electoral calendar	&	342		&	01/07/15\footnote{Page view data made readily available from this date.}-31/12/18		\\
Sports Events       & [January*-December] [2015-18] in sports  &   1,983   &   01/07/15-31/12/18     \\
% Football Matches    & 2017-18 UEFA [Champions, Europa] League  &   330     &   12/09/17-26/05/18\footnotemark[4]    \\
Football Matches    & 2017-18 UEFA [Champions, Europa] League  &   330     &   12/09/17-26/05/18\footnote{Corresponding to the 2017/18 season.}    \\
Film                & [2017, 2018] in film  &   547     &   01/01/17-31/12/18     \\   %  278 (2017), 269 (2018)
Holiday             &  Public holidays in [the United States, the United Kingdom, Australia] &    71     &   01/01/18-31/12/18 \\   %   48 + 23
\end{tabular}
\caption{Statistics of Wikipedia data.}
\label{tab:DataStat}
\end{table*}

We downloaded Wikipedia data dumps for hourly page views \citep{wikimediadumps}. For each article associated to each event, we collected hourly page views for 10 days before and after the event day (21 days total). Page views towards Wikipedia redirect pages for the articles in question were also included\footnote{See the following for more details \url{wikitech.wikimedia.org/wiki/Analytics/Data_Lake/Traffic/pageviews/Redirects}}. 
Whilst the summary articles where the initial events are scraped from act as an important traffic-shaping navigational tool for those browsing within Wikipedia, the majority of attention towards the articles studied comes from sources external to Wikipedia, most commonly Google search ~\citep{dimitrov2018query,gildersleve2018inspiration}. As such, we are confident the results are not determined by Wikipedia-specific navigational constraints from our choice of summary page and are representative of wider interest in events from direct information demand. 
%  Data pre-processing
We focus on planned events lasting no more than 1 day and define the peak value as the maximal page view count in a hour within 48 hours from 0:00 UTC on the supplied date of the event. We selected popular events whose peak value is more than 100 views/hour. In this way, 842 events from the initial database were selected for the following analysis: 92, 213, 250, 229, and 58 events for election, sports events, football matches, film, and holiday, respectively.

% \footnotetext[3]{Page view data made readily available from this date.}
% \footnotetext[4]{Corresponding to the 2017/18 season.}
% \addtocounter{footnote}{2}
%\clearpage
%
%
\section{Modeling Anticipation and Response}
\subsection{Model Description}
%  Model, Fit, Insight from the modeling. 
Here, we propose a simple model for the dynamics of collective anticipation and response, i.e., the number of hourly page views of a Wikipedia article before/after the associated event: 
\begin{equation}
    f_{\rm peak}(t) = C(t) D_{\rm peak}(t),     \label{eq:proposed}
\end{equation}
where $C(t)= 1+ \alpha_c \cos( \omega (t-t_c) )$\ $(\omega= 2 \pi/T)$ describes the circadian rhythm, $\alpha_c\ (0 \leq \alpha_c < 1)$ is the amplitude, $T= 1$ (day) is the period, and $t_c$ is the peak time in the daily oscillation. 
The function $D_{\rm peak}(t)$ describes the anticipation and the response to an event: 
\begin{equation}
	D_{\rm peak}(t)= 
	\begin{cases}
		a_- e^{ (t-t_p)/\tau_-} + b_-  \quad (t< t_p) 	\\
		a_+ e^{ - (t-t_p)/\tau_+} + b_+ \quad (t> t_p) 
	\end{cases},
\end{equation}
where $t_p$ is the peak time, $\tau_-$ ($\tau_+$) is the time constant of the anticipation before the peak (the response after the peak),  $a_- (a_+)$ represents the amplitudes, and $b_- (b_+)$ represents the baseline activity before (after) the peak. Note that this model describes activity except for the peak time $t_p$ (hour). 

The proposed model incorporates the essential features of the peaks in Wikipedia: 1) anticipation, 2) response, and 3) circadian rhythm. 
Several studies have attempted to model burst activity on the web and in social media \citep{crane2008,Matsubara2012,Tsytsarau2014,Kobayashi2016,Proskurnia2017,Rizoiu2017}. 
However, to the best of our knowledge, the proposed model is the first to incorporate all three features. 
While we consider the gradual growth of popularity before a peak as anticipation and describe it using an exponential function, this growth can be attributable to endogenous factors such as word of mouth such as in \citet{crane2008}. In this endogenous model, the peak is described as a symmetric power-law function before and after the peak $\propto |t-t_p|^{-\gamma}$. 
Some of existing models incorporate circadian rhythm \citep{Matsubara2012,Kobayashi2016}; however, these models focus on the information cascade after the event and do not consider activity before the peak. 
\begin{figure}[t]
\centering
   \includegraphics[width=10cm]{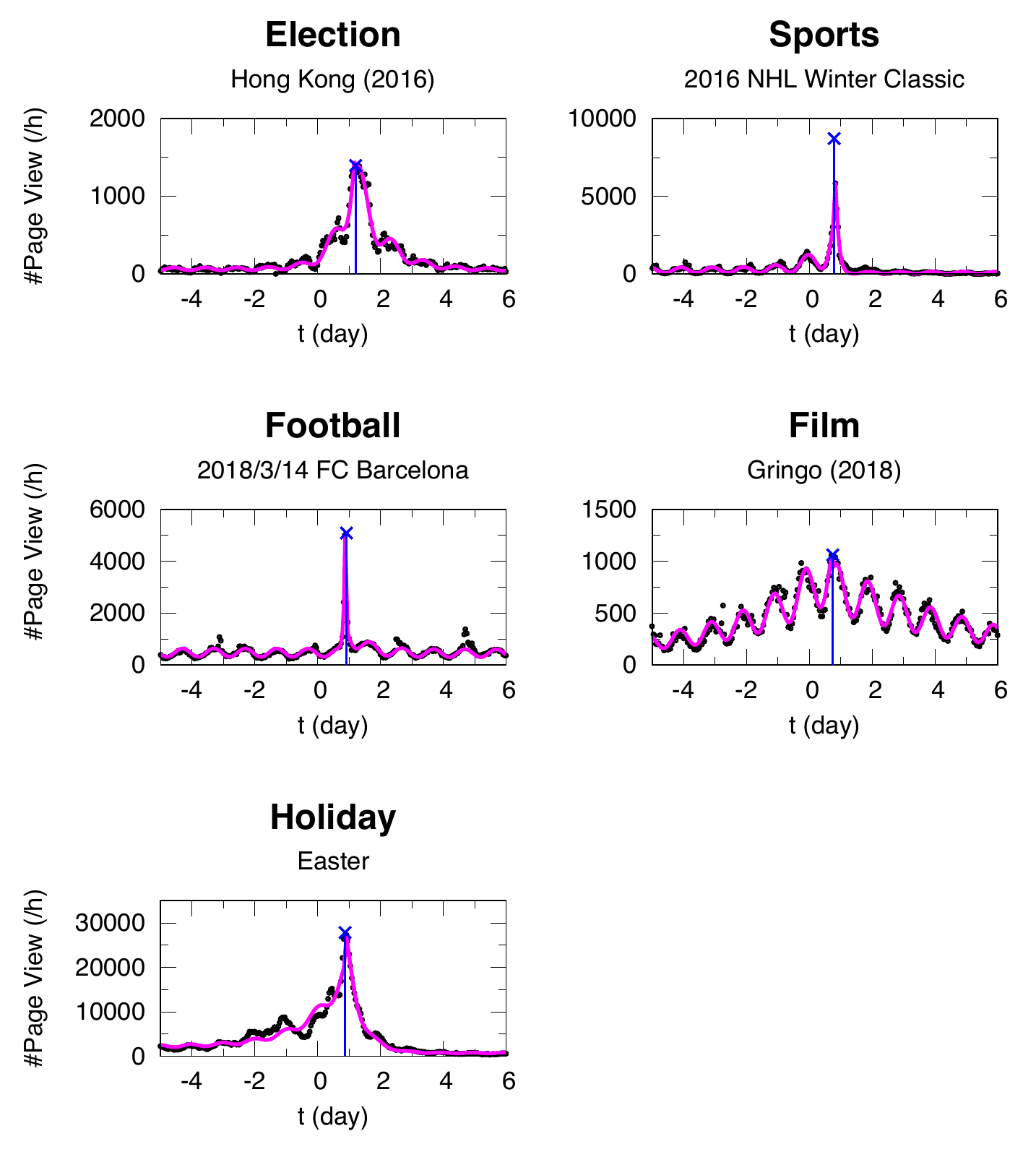}
\caption{The proposed model reproduces collective attention dynamics to a variety of planned events. Examples of five event categories (election, sports, football matches, film releases, and holidays) are shown. 
Data points are shown as dots, the peaks are shown as blue crosses, and the fit by the proposed model is shown in magenta lines.} 
\label{fig:Fit_Main}
\end{figure}
\subsection{Fitting Accuracy} 
\begin{figure}[t]
\centering
   \includegraphics[width=10cm]{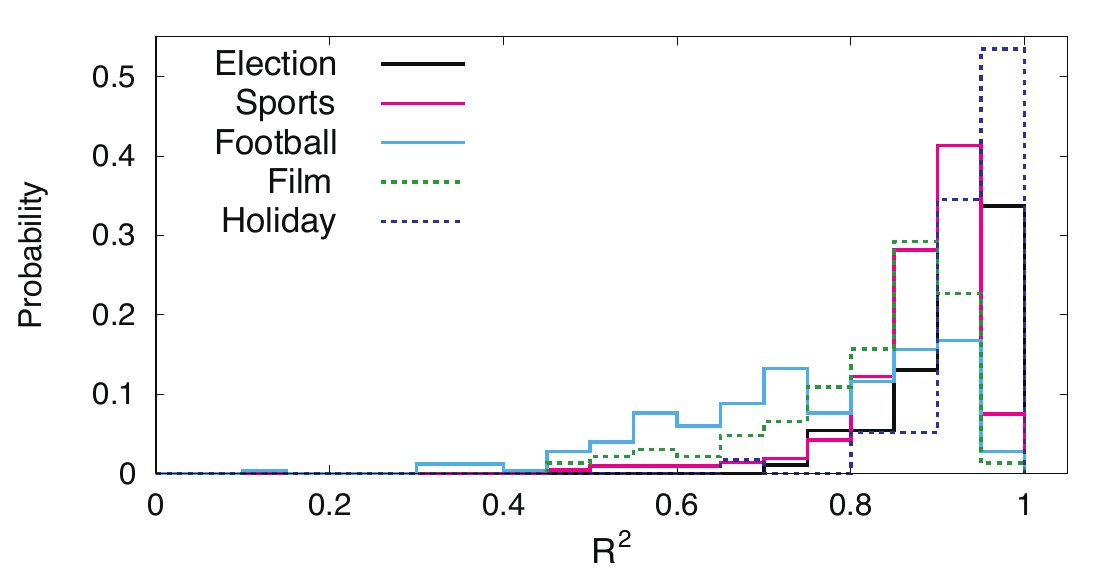}
\caption{
Histogram of Fitting Accuracy. The coefficient of variation $R^2$ was compared among five categories: election (black), sports (magenta), football (cyan), film (dashed green), and holiday (dashed blue).}
\label{fig:Fit_R2}
\end{figure}

We examine whether the proposed model can fit the observed dynamics of collective attention. 
Here, we analyze Wikipedia data related to planned events from five categories (election, sports events, film releases, football matches, and holidays).  
Figure~\ref{fig:Fit_Main} shows that the model accurately fits the temporal patterns in the Wikipedia page view activity for a variety of events. 
The quality of the model fit is evaluated by the coefficient of determination, defined as 
\begin{equation}
	R^2= 1- \frac{\sum_t (s_t- f_{\rm peak}(t))^2}{\sum_t (s_t- \langle s \rangle)^2},
\end{equation}
where $s_t$ is the time series of page view data, i.e., the number of page views in the last hour, 
$\langle s \rangle$ is its average over time, and the summation is over all time points except for the peak time $t_p$. 
The coefficient $R^2$ measures the amount of variance described by the model, and is high when the model can fit the data accurately. 
Note that the maximum value of 1 achieved is only if the model perfectly fits the data: $s_t= f_{\rm peak}(t)$ for any time $t$. 
The model accurately fits the Wikipedia page view data: 
median %(interquartile range) 
of the coefficient $R^2$ was 0.88, %(0.78$-$0.92), 
which indicates that our model describes 88\% of the variation in data. 
Figure~\ref{fig:Fit_R2} shows the histogram of the coefficient $R^2$ for each category.
While the model can fit the peaks of election, sports, holiday  events very accurately, the case of football events is more delicate. This loss of accuracy could originate from the fact that teams may play another match in a different competition within several days of the studied event, inducing additional peaks. 
In addition, we compared the fitting accuracy of the proposed model to that of two existing methods, SpikeM \citep{Matsubara2012} and power-law model \citep{crane2008}. The fitting accuracy of the  proposed method was better than both of these methods, with median % (interquartile range) 
$R^2$ of 0.76 %(0.58$-$0.86) 
and 0.74 % (0.55$-$0.87) 
for SpikeM and the power-law model, respectively. 
\subsection{Interpretability of Model Parameters} 

%  Response parameters  
\begin{table}[t]
\centering
\begin{tabular}{l|ccc}
            &   $\tau_-$ (hours)       &   $\tau_+$ (hours)      &   $\rho$     \\  \hline
Election    &   6.2 (3.7$-$14)      &   19  (14$-$25)   &   0.5 (0.4$-$0.8)    \\
Sports      &   7.0 (2.5$-$18)      &   14  (11$-$17)   &   0.7 (0.4$-$1.0)   \\
Football    &   1.1 (0.8$-$5.3)     &  6.7  (1.3$-$11)  &   0.9 (0.8$-$1.0)    \\
Film    &       34  (20$-$56)       &   87  (55$-$140)  &   0.7 (0.6$-$0.8)    \\
Holiday    &   11 (7.7$-$17)        &   12  (9.0$-$16)   &   1.3 (1.1$-$1.7)    \\
\end{tabular}
\caption{Fitted parameter distribution (Median and interquartile ranges): Anticipation and response time constants $\{ \tau_-, \tau_+ \}$ and the ratio $\rho = S_- / S_+$. }
\label{tab:Par_Response}
\end{table}

The proposed model has three types of parameters: anticipation $\{ a_-, b_-, \tau_- \}$, response $\{ a_+, b_+, \tau_+ \}$, and circadian rhythm $\{ \alpha_c, t_c \}$. $a, b$, and $\tau$ represent the amplitude, baseline views, and time constant for anticipation and response, and $\alpha_c, t_c$ represent the circadian amplitude and offset. The time constants depend on the event category (Table \ref{tab:Par_Response}). For example, the anticipation and response of a film release tend to be slow ($> 1$ day) compared to the other categories. In contrast, a football match demonstrates small time constants ($\leq 7$ hours). Viewers typically lose interest in the next morning. In addition, we quantify the anticipation-response ratio as $\rho= S_-/S_+$, where $S_-= \int_{-M}^0 D_{\rm peak}(s) ds$, $S_+= \int_0^M D_{\rm peak}(s) ds$ (the area under the anticipation / response curve), and $M= 7$ (days) is the time window. We observe that most events (except for Holidays) are response dominated (Table \ref{tab:Par_Response}). 

The circadian parameters are associated with the regional (time zone) distribution of viewers. To demonstrate this, we infer the ratio of viewers in the United States (US), United Kingdom (UK), and Australia (AU) time zones towards individual articles. Assuming that most viewers come from the US, UK, or AU, the fitted circadian function $C(t)$ can be decomposed into three waves: 
\begin{eqnarray}
  C(t)= p_{\rm US} C_{\rm US}(t) + p_{\rm UK} C_{\rm UK}(t)  + p_{\rm AU} C_{\rm AU}(t), 
  \label{eq:Decomposition}
\end{eqnarray}
where $C_{\rm X}(t)= 1+ \bar{\alpha} \cos\left( \omega (t-t_X) \right)$ is the circadian function in $X \in \{$US, UK, AU$\}$, and $\bar{\alpha}= 0.9$ is a constant. 
The reference time $t_X$ was estimated by fitting the circadian function $C(t)$ from the page view data of public holidays and calculating the average of the peak time $t_c$ for the domestic holidays, such as Flag Day (United States), Ayr Gold Cup, and Melbourne Cup ($\alpha_c > 0.5$). 
The estimate was $t_X=$ 20.6, 16.2, and 5.9, for US, UK, and AU, respectively, which reflects the time difference among these regions (US: UTC-5, AU: UTC+10). 
The viewer ratio $\{ p_{\rm US}, p_{\rm UK}, p_{\rm AU} \}$ was determined using the least squares method. 
Next, we analyze the Wikipedia data related to the football matches (UEFA Champions league 2017-18) and film releases. The UK dominated the attention of nearly all the football matches (98 \%: 244/250), and the US dominated the attention of most of the films (81 \%: 186/229). These results indicate that Wikipedia page views reflect the popularity in the real world; all football teams in the UEFA Champions league are based in Europe and the 2018 box office revenue of the US (11.9 billion US dollars) is much greater than that of UK and AU (1.7 and 0.8 billion US dollars, respectively)~\footnote{\url{www.boxofficemojo.com/year/?area=USA},  \\
\url{www.boxofficemojo.com/year/?area=UK},  \\ 
\url{www.boxofficemojo.com/year/?area=AU}}. 
\clearpage
\section{Experimental Evaluation}
We present two applications of the proposed model: 
\begin{itemize}
   \item  predicting the number of page views after the peak;
   \item  clustering page view time series;
\end{itemize}
Based on the proposed model, we develop a forecasting method and implement a clustering method. Then, these methods are applied to the Wikipedia data set, and their performance is compared to the existing methods. 
\subsection{Predicting the Number of Page Views} 
Here we formally define the prediction problem. Given the page view time series $s_t$ up to the observation time $t_{\rm obs}$, we seek to predict the page view activity during prediction period $t_{\rm obs} < t \leq M (= 7$ days). 
Figure~\ref{fig:Pred_Example} shows examples of the prediction results.
\subsubsection{Bayesian Method for Predicting Page View Time Series} 
We develop a method for predicting the future view activity of a Wikipedia article based on the proposed model. 
First, the circadian parameters $\{ \alpha_c, t_c \}$ and anticipation parameters $\{ a_-, b_-, \tau_- \}$ are determined by minimizing the least squared error before the peak. 

%  Bayes Method
We then employ a Bayesian approach for fitting the remaining parameters (response parameters), $\vec{\theta}_+= \{ a_+, b_+, \tau_+ \}$, from short observations. 
The Gaussian distribution is assumed for the observed data 
\begin{equation}
	P \left(s_t| \vec{\theta}_+ \right)= N( s_t |f_{\rm peak}(t), \sigma_n^2), 
\end{equation}
where $N(x| \mu, \sigma^2)$ is the normal distribution with the mean $\mu$ and the variance $\sigma^2$. 
The log-normal prior distribution is assumed to utilize the categorical information (e.g., election) and the page view activity before the peak. 
\begin{equation}
	P \left( \vec{\theta}_+ \right)= \prod_{q_+ \in \vec{\theta}_+} {\rm LN}( q_+| c_{q, k} \hat{q}_-+ d_{q, k}, \sigma^2_{q, k}),  
\end{equation}
where $q_+ \in \{ a_+, b_+, \tau_+ \}$  is a response parameter and $\hat{q}_- \in \{ \hat{a}_-, \hat{b}_-, \hat{\tau}_- \}$ is the fitted value before the peak, and $k$ represents the event category. 
${\rm LN}(x| \mu, \sigma^2)= \frac{1}{\sqrt{2 \pi} \sigma x}  e^{- \frac{(\log x- \mu)^2}{2  \sigma^2} }$ denotes the log-normal distribution. Hyper-parameters $\{ c_{q, k}, d_{q, k} \}$ were determined from fitted parameters. 
The response parameters are determined by maximizing the posterior probability 
\begin{equation}
   P \left(\vec{\theta}_+ | \{s_t \} \right) \propto  P \left( \{s_t \} | \vec{\theta}_+ \right) 	P \left( \vec{\theta}_+ \right).
\end{equation}
Note that the Bayesian method is equivalent to the least squares method when we assume uniform prior distribution: $P \left( \vec{\theta}_+ \right) \propto 1$.
Finally, the page view activity is predicted by calculating the model (Eq. \ref{eq:proposed}) using the fitted parameters. 
\subsubsection{Evaluation metrics} 
Absolute Percentage Error (APE) of page view  time series is used to evaluate the prediction error, ${\rm APE}= \sum_{t=t_{\rm obs}+1 }^M \frac{|s_t- \hat{s}_t|}{N}$, where $N= \sum_{t=t_{\rm obs}+1 }^M s_t$ is the total number of page view during the prediction period, $\hat{x}$ denotes the predicted value of $x$. 
% In addition, we calculate the prediction error of the cumulative page view count (Appendix B).
%
%
\begin{table}[t]
\centering
\begin{tabular}{l|ccc}
Prior                   &   1 day       &   2 days      &   3 days  \\  \hline
No prior                &    0.71       &   0.57        &   0.49    \\
Anticipation            &    0.68       &   0.58        &   0.50    \\
Anticipation, Category  &   {\bf 0.54}  &   {\bf 0.51}  &   {\bf 0.46}   \\
\end{tabular}
\caption{
Prediction error of the Bayesian method. 
Three methods are compared: No prior, Anticipation (only anticipation parameters are utilized, i.e., hyper parameters $c_{q, k}, d_{q, k}$ do not depend on the category $k$), and Anticipation, Category (anticipation parameters and event category are utilized).
The best method is shown in bold. 
The observation time $t_{\rm obs}$ is varied from 1 day to 3 days.  
}
\label{tab:Pred_TS_Bayes}
\end{table}
\subsubsection{Prediction results} 
We evaluate prediction performance by analyzing the Wikipedia data set, and compare the performance of the proposed method to state-of-art approaches.

%  1) Category information can improve prediction 
% \textcolor{blue}{
We develop a Bayesian method for fitting the response parameters $\{ a_+, b_+, \tau_+ \}$, which utilizes the attention dynamics before the peak and event category information. 
First, we evaluate the prediction performance of the Bayesian methods (Table \ref{tab:Pred_TS_Bayes}). 
Incorporating both anticipation parameters and category information substantially improved the accuracy for short observations.
In contrast, the utilization of anticipation parameters only did not improve the accuracy. 
% } 
This result indicates that the event category information supplements the insufficiencies of the page view data.

%
%
%  Prediction performance: Time series. (Table 4)  
\begin{table}[t]
\centering
\begin{tabular}{l|cccc}
Method          &    Complexity     &   1 day       &   2 days      &   3 days     \\  \hline
Proposed        &    8              &   {\bf 0.54}  &   {\bf 0.51}  &   {\bf 0.46}     \\
% Proposed: LS        &    8              &   0.65       &   0.47        &   0.35      \\
SpikeM          &    7              &   0.81        &   0.68        &   0.59    \\
Power-law       &    4              &   0.68        &   0.64        &   0.62    \\
LR              &    $\geq$ 192     &   0.69        &   0.56        &   0.54    \\
LSTM            &    $\approx$ 160,000     &   1.4         &   1.0         &   0.83    \\
\end{tabular}
\caption{Comparison of prediction errors among the proposed method and existing methods. The best method is shown in bold. 
Complexity represents the number of parameters of each method to describe a peak. 
} 
\label{tab:Pred_TS}
\end{table}

% 2) Example of prediction: Fig.~\ref{fig:Pred_Example}.
% 3) Our method predict more accurately than existing methods: Table 4. 
Next, we compare the performance of the proposed method to four existing methods:  SpikeM \citep{Matsubara2012}, Power-law model \citep{crane2008,Tsytsarau2014}, Linear Regression (LR) \citep{Szabo2010}, and LSTM \citep{Hochreiter1997,Mishra2018}. 
For details see appendix A.  
Figure~\ref{fig:Pred_Example} shows examples of time series prediction, which demonstrates that the proposed method outperforms the baseline methods. Though SpikeM reproduces the circadian rhythm, it tends to overestimate the page view activity. 
Furthermore, the proposed method provides the most accurate predictions in terms of the time series (Table \ref{tab:Pred_TS}). 
We observe an improvement of around 20\% over the two runners up (Power-law and LR) for time-series prediction from 1 day observations. 
% \textcolor{blue}{
Furthermore, we confirm that the result is qualitatively the same for the prediction performance of the cumulative view count (Appendix B). % }
%  The median APE was 0.55, 0.68, 0.69, for the proposed method, power-law model, and LR, respectively. 
%
%
\begin{figure}[t]
\centering
   \includegraphics[width=10cm]{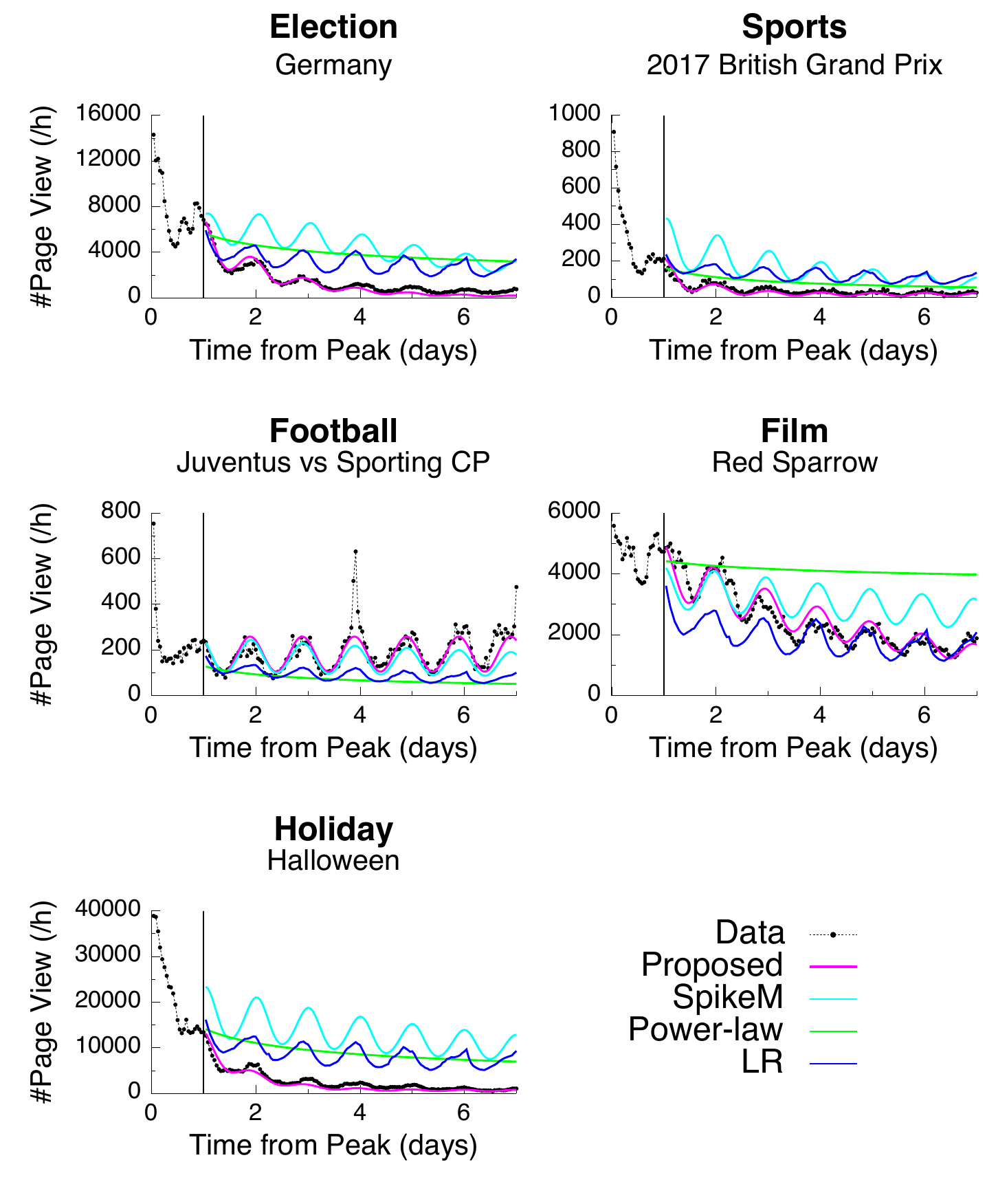}
\caption{Predicting future page view activity. The proposed method (magenta) outperforms existing methods. Vertical bar represents the observation time $t_{\rm obs}= 1$ (day). 
Results from the LSTM approach are omitted from the figure as its predictions often lie outside a reasonable plot range.
} 
\label{fig:Pred_Example}
\end{figure}
 \clearpage
\subsection{Clustering Analysis}
We implement a method for clustering time series data based on the proposed model (Eq. \ref{eq:proposed}). 
The anticipation and response parameters $(a_-, a_+; \tau_-, \tau_+; b_-, b_+)$ are $\log$-transformed before the clustering because they exhibit heavy-tailed distributions. The transformed parameters, along with the circadian parameters $(\alpha_c, t_c)$, are then used as features in clustering with a Gaussian Mixture Model \citep{reynolds2009gaussian}.
The number of clusters is determined based on the Bayesian Information Criterion (BIC) \citep{schwarz1978estimating}, which yields six clusters. The clusters are robust with respect to the initial conditions. 

\begin{figure}[t]
\centering
  \includegraphics[width=10cm]{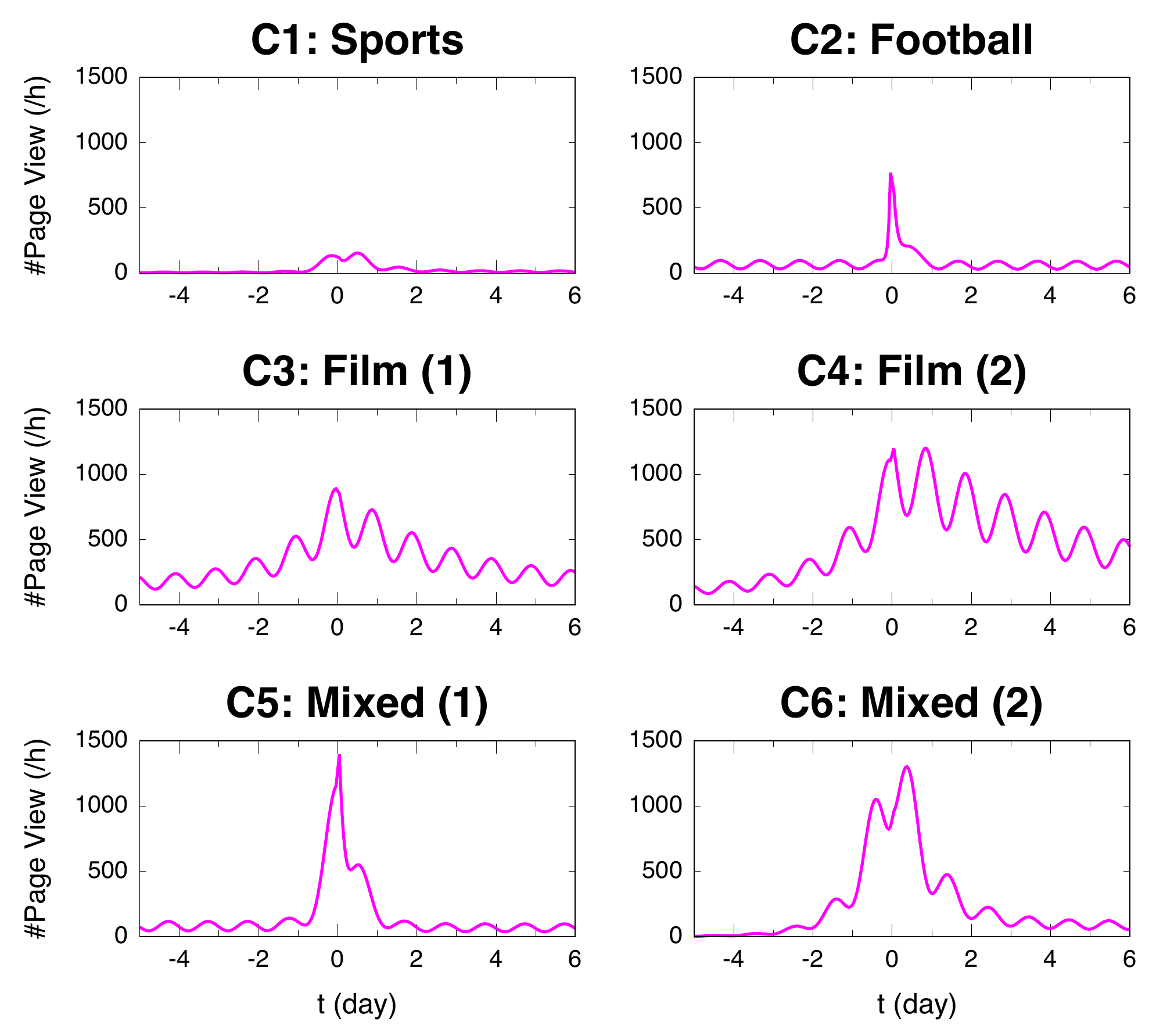}
\caption{Identified cluster centers based on the parameters of the proposed model. The cluster ID and its corresponding category are shown in top of each panel.}
\label{fig:clust_Peak}
\end{figure}

Figure~\ref{fig:clust_Peak} shows the centers for $K=6$ clusters, and Table~\ref{tab:Mclust_model} shows the distribution of each category. 
Four clusters (C1, C2, C3, and C4) correspond to a single category: sports events, football matches, and film release, whereas all the categories are mixed in the other clusters (C5 and C6).  
Cluster C1 exhibits two peaks before/after the main peak due to the circadian rhythm, which corresponds to typical activity for minor sports events.
Cluster C2 exhibits a quick rise and decay, which corresponds to typical activity for football matches. 
The activity for film releases are in clusters C3 and C4, with a slow rise and decay.
Cluster C4 represents the activity for films more popular than C3, decaying more slowly.    
The other two clusters contains all event categories, and cluster C6 shows a slower pattern than C5.

\begin{table}[t]
\centering
\begin{tabular}{l|cccccc}
         & C1  & C2  & C3  & C4 & C5 & C6 \\   \hline
Election & 3   & 2   & 4   & 3  & 16 & 64 \\
Sports   & {\bf 120} & 7   & 1   & 4  & 11 & 70 \\
Football & 0   & {\bf 165} & 0   & 0  & 5  & 63 \\
Film     & 0   & 0   & {\bf 140} & {\bf 57} & 26 & 6  \\
Holiday  & 0   & 0   & 0   & 0  & 14 & 44
\end{tabular}
\caption{Identified clusters (C1, C2, $\cdots$, C6) based on model parameters. Bold indicates that the cluster is dominated by a category ($> 80$\%).}
\label{tab:Mclust_model}
\end{table}

%  Comparison of Clustering Quality
We evaluate the quality of clustering by measuring the similarity to the event category defined by Wikipedia editors. The similarity is quantified by using adjusted mutual information (AMI) \citep{vinh2010information} that takes a value from 0 (independent) to 1 (perfectly correlated), and accounts for increased baseline mutual information between partitions with a larger number of clusters. Here we compare the clustering result from our approach against those obtained using parameter sets of alternative models; SpikeM and the power-law model, as before, as well as the method proposed by \citet{Lehmann2012} which is based on the fraction of total views before, during, and after a peak $\{f_-, f_p, f_+\}$. The previously mentioned linear regression model is not considered, owing to its dependence of the number of parameters on the prediction window, 
% \textcolor{blue}{
likewise the LSTM model with its many parameters. %  } 
We compare against models with the number of clusters that minimise the respective BIC. We also tested alternative models with six clusters, to compare against our own six cluster model, but performance for each of the alternative models was worse than that of their BIC selected clusterings. 
Results are displayed in Table \ref{tab:Cluster_compare} and our approach comfortably outperforms the other models, with median AMI of 0.47 over 10,000 different initial conditions.

% Updated sample tables
\begin{table}[]
\begin{tabular}{l|ccc}
Method     & Clusters & Features & AMI              \\ \hline
Proposed  & 6        & 8        & {\bf 0.47 (0.46-0.49)} \\
SpikeM    & 8        & 7        & 0.35 (0.33-0.38) \\
Power-law & 7        & 4        & 0.36 (0.32-0.38)  \\
Fraction   & 2        & 3        & 0.39 (0.39-0.39) \\
\end{tabular}
\caption{Median and interquartile range for AMI over 10,000 initial conditions across the different models. Fraction means the clustering based on the fractions of total views before, during, and after a peak. The best method is shown in bold.}
\label{tab:Cluster_compare}
\end{table}
\vspace{3cm}
\section{Relating Event Outcome and Collective Attention Dynamics}
The results presented in the previous section suggest that the temporal properties of collective attention (e.g., the parameters of the proposed model) are associated with the event category (e.g., football matches). 
In this section, we investigate the link between more detailed information about the event and the model parameters. Here, we restrict our analysis to a dataset associated with football matches for easy interpretation of the results. 

First, we examine how the model parameters depend on the stage of competition (group stage, knockout stage, and the final match in the Champions League) and the match result (win, draw, and lose). 
As shown in Table~\ref{tab:par_stage}, the anticipation and response amplitude ($a_-$, $a_+$) depend on the stage of competition, with matches in the latter stages of the competition being more popular. The time constants ($\tau_-$, $\tau_+$) do not depend on the stage (data not shown).

\begin{figure}[t]
\centering
  \includegraphics[width=8cm]{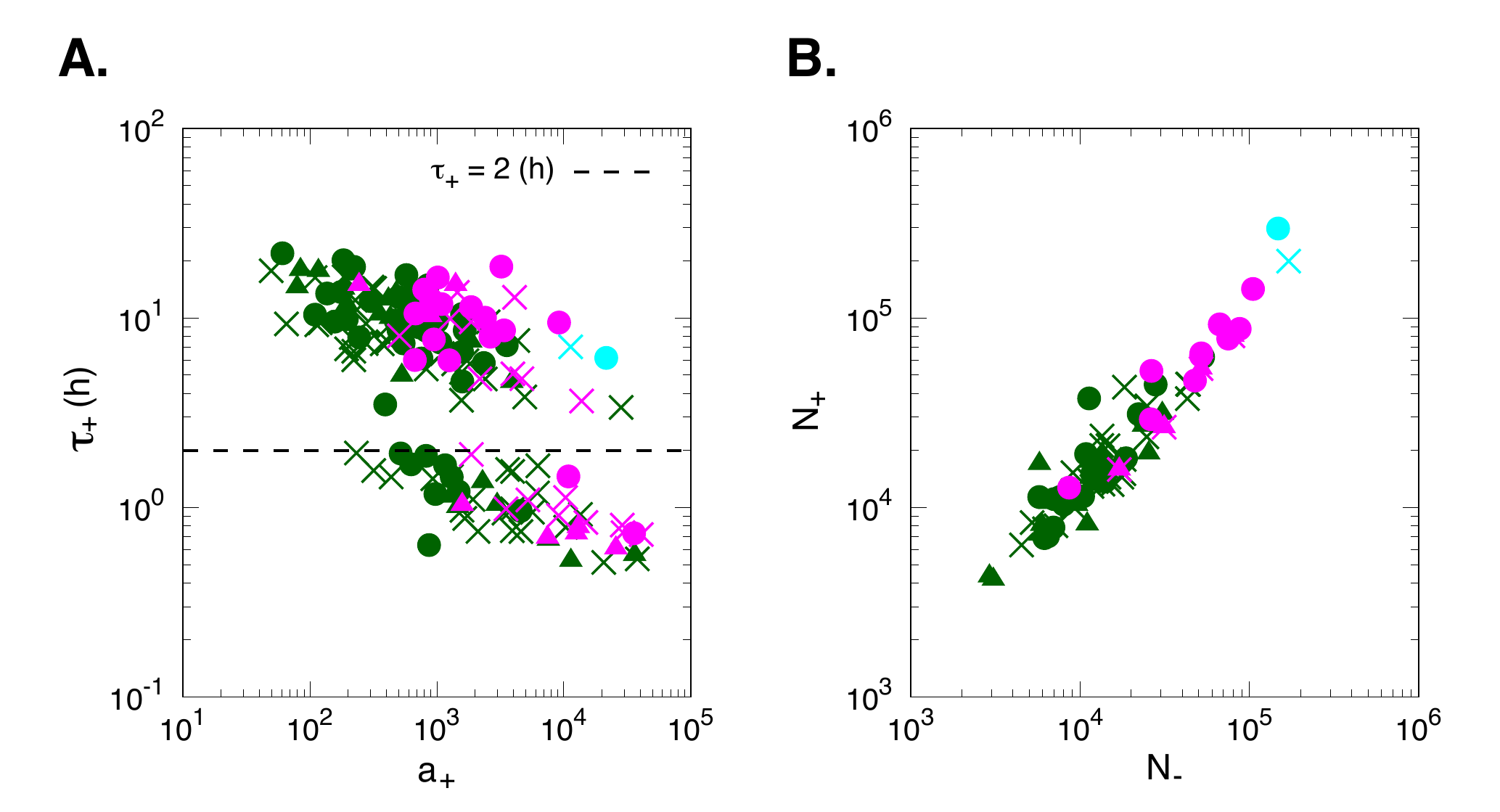}
\caption{Effect of stage and result of a football match on response dynamics (2017-18 UEFA Champions League). 
A. Reaction parameters ($a_+, \tau_+$). The peaks with poor fitting accuracy ($R^2 < 0.7$) are omitted from this figure. 
B. Page view counts before and after the peak ($N_-, N_+$).
Colors represent the stage, i.e., group stage: green, knock Out stage: magenta, final: cyan. Symbols represent the results, i.e., win: circle, draw: triangle, and lose: cross.}
\label{fig:Football-Response}
\end{figure}

\begin{table}[t]
\centering
\scalebox{0.95}{
\begin{tabular}{l|cc}
            &   $a_-$                   &   $a_+$       \\  \hline
Group       &   940  (280$-$2,400)      &   580  (220$-$1,500)     \\  
Knockout    &   5,500 (2,300$-$11,000)  &   2,200 (1,200$-$9,100)     \\
Final       &   27,000                  &   16,000       \\
\end{tabular}}
\caption{Median and interquartile ranges of parameters for different stages (no range for final).} 
\label{tab:par_stage}
\end{table}
Interestingly, the response parameters ($a_+, \tau_+$) of the losing teams are distinct from the winning ones (Table~\ref{tab:par_res}). 
We can observe two clusters in the response parameters that correspond to excited ($\tau_+ > 2$ hours) and disappointed ($\tau_+ <2 $ hours) response (dashed line in Fig.~\ref{fig:Football-Response}A). 
The proportion of losing teams in the disappointed class is much higher than that in the excited one: 55 \% (50/91) and 31 \% (49/159), respectively. Note that the disappointed response cannot be identified based solely on the view counts before and after the peak (Fig.~\ref{fig:Football-Response}B). Ahead of their Champions League semi-final on 2\textsuperscript{nd} May against AS Roma, Liverpool FC manager J\"urgen Klopp warned that ``no one remembers losers'' \cite{bascombe_2018}. Liverpool won the two-legged tie, and our results clearly confirm Klopp's statement - with Liverpool FC's $\tau_+=12.9$ compared to AS Roma's $\tau_+=0.7$ (hours).

\begin{table}[t]
\centering
\begin{tabular}{l|cccc}
        &   $a_+$               &   $\tau_+$ (hours)       \\  \hline
Win     &   680 (250$-$1,300)   &   8.6 (2.5$-$12)   \\
Draw    &   770 (200$-$1,900)   &   9.4 (1.0$-$15)   \\
Lose    &   1,400 (370$-$4,600) &   1.9 (1.0$-$7.9)  \\
\end{tabular}
\caption{Median and interquartile range of response parameters for different match results.}
\label{tab:par_res}
\end{table}
%
%
%  Excited class ($\tau_+ >2$): Win: 23, Draw: 16, and Lose: 19. 
%  Disappointed class ($\tau_+ \leq 2$): Win: 4, Draw: 5, and Lose: 18. 

To further quantify the association between the model parameters and match result, we infer whether a football team wins, draws, or loses a match based on the page view time series of the team and its opponent. 
%  Setting
A linear support vector machine classifier\footnote{Best performance is achieved with a linear kernel, but the results are qualitatively similar for radial basis function, polynomial, and sigmoid kernels.} \citep{Chang2011SVM} implemented in scikit-learn \citep{scikit-learn} is used for the classification task. 
We consider two kinds of features: model parameters (based on the proposed model, SpikeM, and Power-law model) and a feature set based on the fractions of total views before, during, and after a peak \citep{Lehmann2012}. 
The classification performance is evaluated by 5-fold cross-validation and also compared against a baseline of always predicting the most frequent result (draw). 

%  Results
Table~\ref{tab:SVMperformance} summarises the classification performance of the selected parameter combinations against alternative models (other combinations with lower performance or equivalent performance and more features omitted). 
Classification based on the parameters of the proposed model yields better performance than the other approaches. 
The best performance is achieved by the features of the response parameters \{$a_+$, $b_+$, $\tau_+$\} of both the target and opponent teams. 
Note that the other parameters, e.g. the anticipation and the circadian parameters, do not improve the performance.

\begin{table}[]
\centering
\begin{tabular}{l|l|c}
Model       & Features                              &  Accuracy     \\  \hline
Proposed    & \{$a_+$,$b_+$,$\tau_+$\}$^{opp}$      &   {\bf 68\%}  \\
            & \{$a_+$,$b_+$,$\tau_+$\}              &   56\%        \\ 
SpikeM      & \{$u(0)$,$\beta$,$t_b$,$s_b$,$\epsilon_0$,$p_a$,$p_s$\}$^{opp}$ & 57\% \\ 
Power-law   & \{$a_+$,$\gamma_+$\}                  	&   58\%   \\  
Fraction     & \{$f_-$,$f_p$,$f_+$\}                 	&   55\%   \\
Baseline    & -									& 42\% \\
\end{tabular}
\caption{Classification accuracy of football match results.
By default, only the target team's parameters are included, superscript `$opp$' indicates the opposing team's parameters are also included. The best method is shown in bold.}  
\label{tab:SVMperformance}
\end{table}

These results indicate that the parameters of the proposed model are not only able to capture the the different temporal patterns of different event classes, but are also sensitive to detailed event-related information, 
in this case the stage of competition and match result. Finally, we analyzed another series of football matches (2017-18 UEFA Europa League) and the results are qualitatively similar to those of the Champions League data. 
\clearpage
% \vspace{3cm}
%
%
\section{Discussion and Limitations} 
As set out in the introduction, the two objectives of this paper were to investigate the two research questions: 
\begin{description}
    \item[RQ1]	What are the essential characteristics of collective attention dynamics towards planned events?
    \item[RQ2]	How is event-related information (e.g., category and outcome) associated with the dynamics of collective attention?
\end{description}

%	RQ1:    Characterization 
To address RQ1, we have collected hourly time series of page views for a range of topics. This data set is the first to focus on the hourly page viewing activity observed on relevant Wikipedia pages 10 days before and after each planned event, thus allowing to compare the dynamics of the collective attention to various categories of events. 
We have then proposed a simple model for characterizing the time series of collective attention, which has helped reveal some of their essential properties: 
\begin{itemize}
	\item	Attention dynamics exhibits exponential anticipation and response, and circadian rhythm; 
	\item	The time constants of anticipation and response ($\tau_+, \tau_-$) are event-related:  
	for example, time constants are shorter for football matches and longer for film releases; 
	\item	Circadian rhythm is associated with the geographical distribution of attention.  
\end{itemize}
The results on anticipation suggest an exponential distribution of the number of  interested viewers before an event. An exponential increase in page-views can be attributed to the characteristics of independent individuals rather than to network effects (e.g., information diffusion on social network). This is because the time constant depends on the category of the event and it is larger (more than 10 hours) than that of information diffusion. 
Moreover, the technological affordances of Wikipedia as a website are very different from social networks such as Facebook, Twitter, and YouTube, where interactions between users and `trending' items are much more emphasised and lead to network popularity phenomena.  

%	RQ2:	Association between the attention dynamics and the event information 
We have examined RQ2 through prediction analysis and clustering analysis. 
We have shown that the event category and the anticipatory pattern before the peak improve the prediction performance for the response pattern, even for short observation windows (Table 3).  
The clustering results of the attention patterns are akin to the event categories annotated by Wikipedia editors (Table 5).  
These results indicate that the temporal patterns of collective attention are highly concomitant with the event category (e.g., sports versus elections). 
In addition, we performed a classification task of the event outcome (win/draw/lose) using a data set of association football matches (UEFA Champions League 2017-18). We found that the realization of an event drives the dynamics of collective attention and demonstrated that the event outcome can be inferred from collective attention patterns. 

%   Future work: Which type of events exhibit anticipatory pattern? 
These results suggest that events should be classified based on whether their {\it sentimental outcome}  is known in advance or not. We define the concept of sentimental outcome according to how positive/negative a user feels after the event. The distribution of all users' feelings is then representative of the sentimental outcome of the event. Events that elicit positive responses such as popular movies or holidays have positive sentimental outcome, and events with negative responses such as disasters or a team losing a football match have negative sentimental outcome. The sentimental outcome is not always known in advance, and can also be bimodal. For example, the sentimental outcome of a football match is largely contingent on its result: the users will be excited (disappointed) if the team they support wins (loses) the match. This differential response is not observed for movies, though the sentimental outcome could vary depending on movie quality/success. 
Additionally, holidays typically do not have explicit outcome so variation in the response dynamics is not expected. 
This is confirmed by the relative sizes of the interquartile ranges for $\tau_+$ in table \ref{tab:Par_Response}. Unfortunately, due to limitations on the structured Wikipedia data, we are unable to test more generally this hypothesis of differential excited/disappointed responses based on the sentimental outcome, as we have done for the football matches. Our results also suggest that certainty of positive sentimental outcome is associated with the anticipatory activity. 
We found that most of the film releases and holiday events exhibit anticipatory activity, whereas most of election and sports events do not exhibit it (Fig. 1). Future research, using sentiment analysis on user generated content (e.g. data from Twitter \citep{dodds2011temporal} or Reddit \citep{medvedev2017anatomy}), could help investigate which type of events exhibit a strong anticipatory pattern by quantifying the sentimental outcome with tools from natural language processing. 
More generally, we have not fully explored the multitude of possible relations between event content and patterns of collective attention.
Enriching our analysis with natural language processing tools in a combination of sources, including news articles and social media, is a promising research direction. 

%   Limitations
%		Limitation 1:  We do not cover all the events. 
Our model can quantify collective attention patterns of events, and can thus help understand  the mechanisms driving the anticipation and response of an event.
However, there are some limitations in this study. 
Events can be classified based on two dimensions: whether or not they are planned in advance, and whether they are scheduled to be held on a single day or over multiple days. 
This study has focused on a specific type of event, i.e., planned events scheduled for a single day. 
The proposed model is validated for this restricted type of events, but it could be generalised for more general situations.
For instance, the model without anticipation $(a_-=0)$ would be reasonable for unplanned events. 
Typically, events held over multiple days encompass several {\it sub-events}, such as in the Olympic games. 
The collective attention towards such events may be described by the sum of multiple peaks: $\sum_k f_{\rm peak}(t; t_{p, k})$, where $t_{p, k}$ is the time of the $k$-th sub-event. 
Further work is required to investigate how to extend our methodology to multi-day events with unknown sub-events and / or continuous stimuli.  
%
%   Limitation 2: We focus on highly popular events.
Moreover the present study has only considered events with a high popularity, owing to their importance online and offline. 
Although the temporal pattern of less trending events is similar, it is more challenging to determine model parameters and predict the response patterns for such events because of the scarcity of data points. 
Point process models~\citep{valera2015,Kobayashi2016} could be suitable for such situations. 
Overall, future studies would thus be required to develop the generalized models and investigate the links between event-related information and attention patterns in a more general setting.  
\section{Conclusion}

In this paper, we have studied collective attention towards Wikipedia pages associated with planned events, defined as events known to happen at a specific date in the future (in contrast to unexpected, unplanned events such as earthquakes or plane crashes). 
As a first step, we have collected hourly time series of page views for a range of topics and proposed 
a simple model by incorporating key features, i.e., the anticipation and the response to an event as well as circadian rhythm. 
To the best of our knowledge, the proposed model for collective attention is unique in taking all of these factors into account.
This model allows us to develop a method for predicting the future time evolution of the popularity and for clustering time series.  
Our methodology outperforms state-of-the-art methods for prediction and clustering tasks, emphasizing the importance of appropriately modeling the dynamics of collective attention. 
Interestingly, the event category information (e.g. football match and film release) improved the prediction accuracy from short observations, 
and the clustering result based on the model parameters was associated with the event categories. 
More specifically, using football match data, we have demonstrated that the response parameters are associated with the type and the result (win, draw, and lose) of the match. 
These results suggest that not only the event category but also that more detailed event-related information drives the temporal pattern of popularity, and led us to postulate the notion of sentimental outcome to explain the differences in attention patterns between events. 
We believe that the proposed model provides an important contribution towards studying the relationship between the dynamics of collective attention and characteristic information underlying an event.
\section*{Acknowledgments}
We thank Kazuhiro Kurita and Yuka Takedomi for stimulating discussions. Furthermore, this paper was greatly improved by the comments of anonymous reviewers. 
R.K. is supported by JSPS KAKENHI Grant Numbers JP17H03279, JP18K11560, and JP19H01133, and JST PRESTO Grant Number JPMJPR1925. 
T.U. is supported by JSPS KAKENHI Grant Numbers JP19H01133, and JST CREST JPMJCR1401. 
\section*{Appendix A: Baseline Models}
We summarize four existing methods for predicting the response dynamics of collective attention.

\begin{itemize}
  \item {\bf SpikeM}~\citep{Matsubara2012}: 
  This model describes the page view count in the the last hour $x(t)$ using a difference equation 
    \begin{eqnarray}
	     && x(t+1) = p(t+1) \nonumber \\
	     && \quad \quad \ \left( u(t) \sum_{k=t_b}^t (x(k)+ s(k) ) f(t+1-k) + \epsilon_0  \right), \nonumber  \\
	     && u(t+1)= u(t)-x(t+1), \nonumber
    \end{eqnarray}
    where $p(t)= 1- 0.5 p_a \{ 1+ \sin( \omega (t+ p_s) ) \}$\ ($\omega= 2 \pi/T$) describes circadian rhythm, $s(t)= s_b$ ($t=t_b$), $s(t)= 0$ (otherwise) represents the shock due to the event, and $f(t)= \beta t^{-1.5}$ is the power-law function. 
    Seven parameters $\{ u(0), \beta; t_b, s_b; \epsilon_0; p_a, p_s \}$ are fitted using the least squares method. This model incorporates the power-law relaxation in response activity and the circadian rhythm in human behavior.
\item {\bf Power-law model}~\citep{crane2008,Tsytsarau2014}:  
  \citet{crane2008} showed that the viewing activity of YouTube videos demonstrates power-law relaxation behavior: 
    \begin{equation}
	    x(t)= 
	    \begin{cases}
		    a_- (t_p-t)^{\gamma_-}  \quad (t< t_p) 	\\
		    a_+ (t-t_p)^{\gamma_+}  \quad (t> t_p) 
    	\end{cases},  \nonumber 
	\end{equation} 
   where $t_p$ is the peak time. Four parameters $\{ a_-, \gamma_-; a_+, \gamma_+ \}$ are fitted by least squares method. 
\item {\bf Linear Regression (LR)}~\citep{Szabo2010}: 
  This method applies linear regression to the logarithm of the cumulative view counts 
    \begin{equation}
	    \log R(t)= \alpha_t + \log R(t_{\rm obs})+ \sigma_t \xi,  \nonumber 
	\end{equation} 
  where $R(t)= \sum_{k=t_p+1}^t x(k)$ is the cumulative view count after the peak, $t_{\rm obs}$ is the observation time, $\xi$ represents Gaussian noise. For each time, the cumulative count is predicted by the unbiased estimator $\hat{R}(t)= R(t_{\rm obs}) \exp(\hat{\alpha}_t + \hat{\sigma}^2_t/2)$, where $\hat{\alpha}_t$ and $\hat{\sigma}^2_t$ are the fitted values obtained using the maximum likelihood method. 
\item {\bf  Long short-term memory (LSTM)}~\citep{Hochreiter1997}: 
  LSTM is a recurrent neural network model for time series forecasting~\citep{Hochreiter1997}. With the advent of deep learning methods, LSTM is also getting popular in social media analysis, including popularity prediction~\citep{Mishra2018} and fake news detection~\citep{Ruchansky2017}. 
  We used the Matlab Deep Learning Toolbox  
  for the implementation and adopted its default parameters~\footnote{\url{https://www.mathworks.com/help/deeplearning/ug/time-series-}
  \url{forecasting-using-deep-learning.html}}. 
  Specifically, the number of hidden units was 200 and the number of model parameters was 160,800. 
\end{itemize}
\clearpage
\section*{Appendix B: Prediction Performance of the Cumulative View Count}
We calculate the Absolute Percentage Error (APE) of cumulative page view count, ${\rm APE}= \frac{|N- \hat{N}|}{N}$, where $\hat{N}$ denotes the predicted cumulative view count after the observation period. 
We evaluate the prediction error of the Bayesian methods (Table~\ref{tab:Pred_Cum_Bayes}). Then, the best Bayesian method was compared to existing methods (Table~\ref{tab:Pred_Cum}). 
The proposed method shows an improvement of around 20 \% over the runner up (LR) for the prediction performance.

\vspace{1cm}
\begin{table}[ht]
\centering
\begin{tabular}{l|ccc}
Prior                   &   1 day       &   2 days      &   3 days      \\  \hline
No prior                &    0.65       &   0.47        &   0.35        \\
Anticipation            &    0.62       &   0.50        &   0.38        \\
Anticipation, Category  &   {\bf 0.44}  &   {\bf 0.41}  &   {\bf 0.33}  \\
\end{tabular}
\caption{
% \textcolor{blue}{
Prediction error of the Bayesian method. 
Three methods are compared: No prior, Anticipation (only anticipation parameters are utilized), and Anticipation, Category (both anticipation parameters and event category are utilized).
The best method is shown in bold. 
The observation time $t_{\rm obs}$ is varied from 1 day to 3 days. % }
}
\label{tab:Pred_Cum_Bayes}
\end{table}
%
%  Prediction performance: Cumulative count. (Table 10)  
\begin{table}[ht]
%\vspace{0.4cm}
\centering
\begin{tabular}{l|cccc}
Method          &    Complexity     &   1 day      &   2 days      &   3 days     \\  \hline
Proposed        &    8              &  {\bf 0.44}  &   {\bf 0.41}  &   {\bf 0.33}    \\
% Proposed: LS        &    8              &   0.65       &   0.47        &   0.35      \\
SpikeM          &    7              &   0.74        &   0.60        &   0.51    \\
Power-law       &    4              &   0.62        &   0.57        &   0.54    \\
LR              &    $\geq$ 192     &   0.54        &   0.51        &   0.50    \\
LSTM            &    $\approx$ 160,000     &   1.3        &   0.90        &   0.74    \\
\end{tabular}
\caption{Comparison of prediction errors among the proposed method and existing methods. The best method is shown in bold. 
% \textcolor{blue}{
Complexity represents the number of parameters of each method. % } 
}  
\label{tab:Pred_Cum}
\end{table}

\bibliography{wikipedia}

\end{document}